\documentclass[aip, jmp, prb, amsmath, amssymb, reprint]{revtex4-1}

\usepackage{graphicx}
\usepackage{dcolumn}
\usepackage{bm}
\usepackage{hyperref}

\usepackage{xcolor}

\usepackage{subcaption}
\newcommand{\labelphantom}[1]{%
{\phantomsubcaption%
\label{#1}}%
}%


\begin{document}

\title{Perturbative Determination of Plasma Microinstabilities in Tokamaks}

\author{A. O. Nelson}
\affiliation{Princeton University, Princeton, NJ 08544, USA}
 
\author{F. M. Laggner}
\affiliation{Princeton Plasma Physics Laboratory, Princeton, New Jersey 08540, USA}

\author{A. Diallo}
\affiliation{Princeton Plasma Physics Laboratory, Princeton, New Jersey 08540, USA}

\author{Z. A. Xing}
\affiliation{Princeton University, Princeton, NJ 08544, USA}


\author{D. R. Smith}
\affiliation{University of Wisconsin-Madison, Madison, Wisconsin 53706, USA}

\author{E. Kolemen}
\affiliation{Princeton University, Princeton, NJ 08544, USA}
\affiliation{Princeton Plasma Physics Laboratory, Princeton, New Jersey 08540, USA}

\date{\today}

\begin{abstract}
Recently, theoretical analysis has identified plasma microinstabilities as the primary mechanism responsible for anomalous heat transport in tokamaks. In particular, the microtearing mode (MTM) has been credited with the production of intense electron heat fluxes, most notably through a thin self-organized boundary layer called the pedestal. Here we exploit a novel, time-dependent analysis to compile explicit experimental evidence that MTMs are active in the pedestal region. The expected frequency of pedestal MTMs, calculated as a function of time from plasma profile measurements, is shown in a dedicated experiment to be in excellent agreement with observed magnetic turbulence fluctuations. Further, fast perturbations of the plasma equilibrium are introduced to decouple the instability drive and resonant location, providing a compelling validation of the analytical model. This analysis offers strong evidence of edge MTMs, validating the existing theoretical work and highlighting the important role of MTMs in regulating electron heat flow in tokamaks. 
\end{abstract}

\maketitle

Utilizing tokamak reactors \cite{Wesson2004} to realize magnetic confinement fusion holds the prospect of producing clean and sustainable energy \cite{Tuck1971,Artsimovich1972}. This effort requires the establishment of hot, dense plasma cores through a self-organized high-confinement regime (H-mode) characterized by steep plasma gradients in a thin region called the pedestal \cite{Wagner1984}. While only covering $\sim10\%$ of the plasma radius, the pedestal can be responsible for up to $\sim70\%$ of the total plasma pressure and fusion performance and is thus essential for the successful optimization of tokamak devices.

A standard H-mode pedestal is characterized by two competing physics phenomena. First, strong velocity shear caused by variation in the radial electric field suppresses transport in the pedestal by tearing apart turbulent eddies, allowing for the formation of steep temperature and density gradients that would otherwise be eliminated by diffusion \cite{Burrell1997}. Second, various metastable microinstabilities induce transport across the pedestal despite the high levels of turbulent shear, controlling the evolution of the pedestal structure \cite{Diallo2021}. If left unmitigated, non-linear interactions between these microinstabilities periodically spark global explosive events \cite{Diallo2018, Lee2017} called edge-localized modes (ELMs) \cite{Connor1998} which can melt and erode the machine wall \cite{Ham2020}. As such, understanding the details of these microinstabilities is not only for crucial for the optimization of fusion parameters but also for successful plasma control \cite{Evans2006, Park2018}.

Over the past few decades, largely theoretical and computational work has uncovered five plasma instabilities that may contribute to inter-ELM transport through the H-mode pedestal. These include three electrostatic modes: the trapped electron mode (TEM), the electron temperature gradient (ETG) mode, and the ion temperature gradient (ITG) mode; and two electromagnetic modes: the kinetic ballooning mode (KBM) and the microtearing mode (MTM). Extensive modeling has shown that each of these modes could become unstable in the tokamak edge under certain conditions, but an \textit{experimental} validation of which modes are actually active in the pedestal remains elusive due to the nebulous nature of the turbulence. Without an empirical determination of individual modes, it is difficult to improve the physics basis of leading turbulent models.

Notably, recent theoretical work suggests that the MTM \cite{Hazeltine1975, Drake1977}, a small-scale resistive magnetohydrodynamic (MHD) mode not yet included in leading predictive models \cite{Snyder2011}, might play a critical role in limiting electron thermal transport through the pedestal \cite{Dickinson2012, Hatch2016}. The presence of pedestal MTMs has been suggested through analysis of so-called ``transport fingerprints" \cite{Kotschenreuther2019} and through comparisons of measured magnetic fluctuations with sensitive theory-based (gyrokinetic) simulations \cite{ Applegate2007,Doerk2011,Guttenfelder2011,Dickinson2012, Dickinson2013,Swamy2014,Hatch2016,Chowdhury2016,Kotschenreuther2019,Hatch2020,Chen2020}. However, a conclusive experimental identification of these modes has not yet been presented and is needed to validate the theoretical results.

In this article, we introduce novel experimental evidence to unambiguously demonstrate the existence of MTMs in the tokamak pedestal. MTMs are theorized to destabilize at particular resonant locations within the plasma and to oscillate at the electron diamagnetic frequency, which is a function of plasma radius. We utilize an innovative experimental technique in the form of large vertical plasma displacements to dissociate these two phenomena by dynamically shifting magnetic surfaces in the edge region. Experimental observations of MTM evolution in a series of plasma discharges on the DIII-D tokamak are found to be in overall agreement with theoretical expectations, providing a compelling validation of the model. The presented work describes a clear-cut experimental identification of MTMs, focusing attention on the need to include MTM physics into predictive tokamak models. 


\section*{Time-dependent MTM identification} 

\begin{figure}
    \includegraphics[width=1\linewidth]{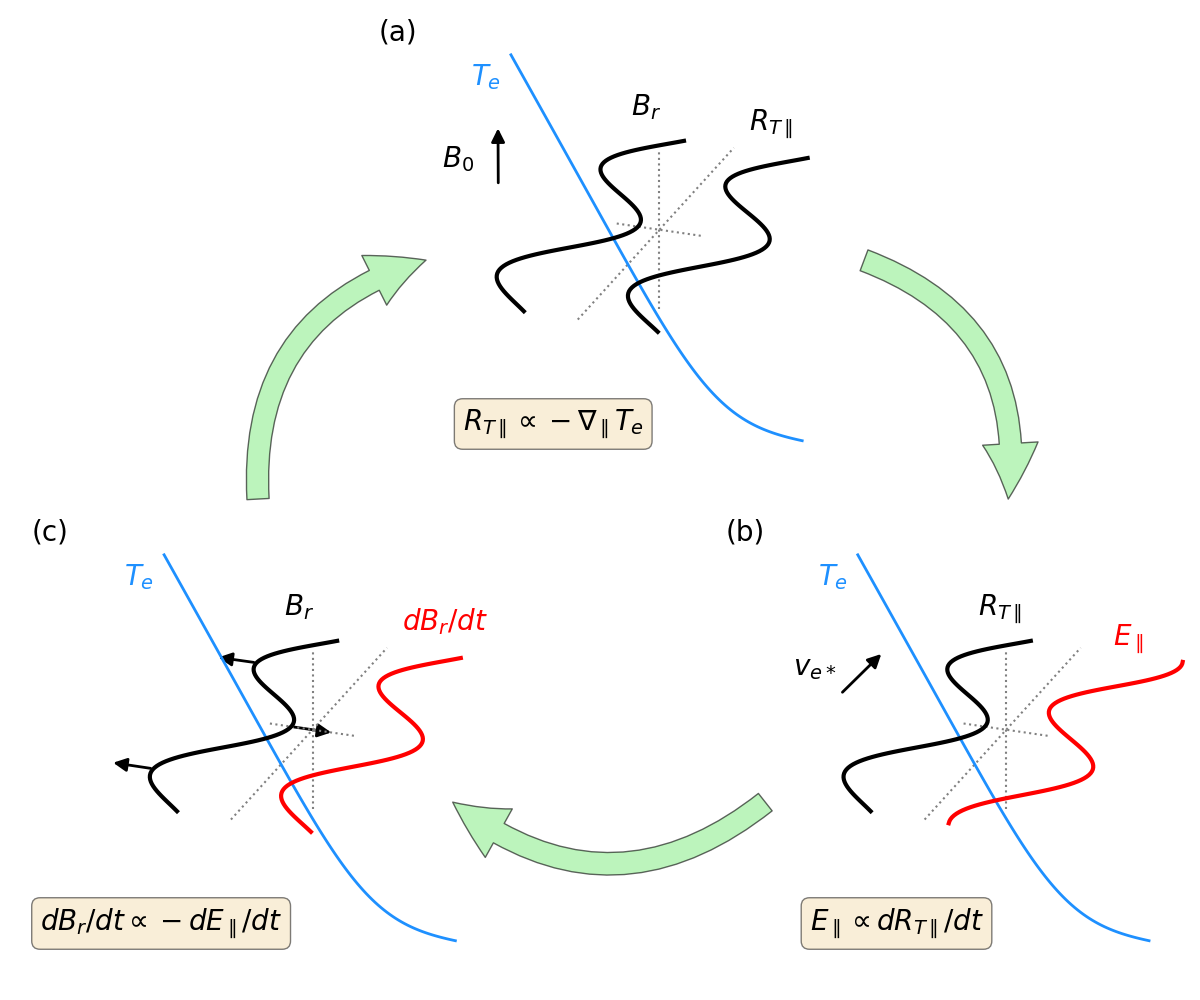}
    \labelphantom{fig:MTM-a}
	\labelphantom{fig:MTM-b}
	\labelphantom{fig:MTM-c}
    \caption{Physics of the time-dependent thermal force drive for MTMs. (a) A temperature gradient $\nabla T_\mathrm{e}$ projected onto a $q$-resonant magnetic perturbation creates spatial variation in the thermal drag force ($R_\mathrm{T\parallel}$) between electrons and ions. (b) Due to parallel motion at $v_\mathrm{e*}$, a time lag is introduced to the thermal force, creating a de-phased electric field through charge separation. (c) The resulting inductive field $dB_\mathrm{r}/dt$ is in-phase with the initial perturbation, causing the instability to grow.}
    \label{fig:MTM}
\end{figure}

\begin{figure*}
    \includegraphics[width=1\linewidth]{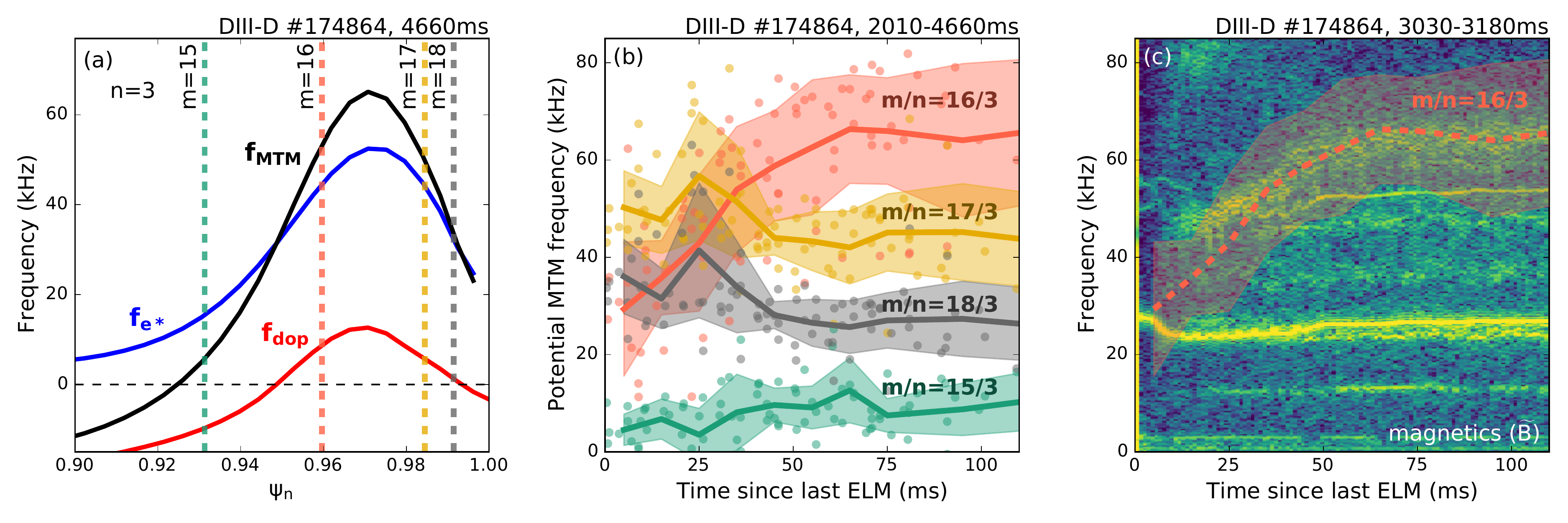}
    \labelphantom{fig:normal-a}
	\labelphantom{fig:normal-b}
	\labelphantom{fig:normal-c}
    \caption{Time-dependent MTM frequencies computed from experimental profiles match well with those observed in magnetic fluctuations. (a) The MTM frequency $f_\mathrm{MTM}$ is shown as the sum of $f_{dop}$ and $f_{e*}$ for a representative timeslice, along with the possible rational $q$ surface locations for an $n=3$ mode. (b) At each rational $q$ surface, the expected MTM frequency is plotted as a function of time since the last ELM. (c) Profile-based predictions for the $m/n=16/3$ mode match exceptionally with the $n=3$ chirped mode observed in magnetic fluctuations.}
    \label{fig:normal}
\end{figure*}

Microtearing modes 
are finite-collisionality electromagnetic modes destabilized by the electron temperature gradient $\nabla T_\mathrm{e}$  \cite{Hazeltine1975, Drake1977}. In tokamaks, magnetic surfaces have a helical structure defined by the ratio $q=m/n$, which describes the winding of a field line in the poloidal ($m$) and toroidal ($n$) directions. At rational values of $m$ and $n$, a radial perturbation $B_{r}$ can be driven unstable if the presence of $\nabla T_\mathrm{e}$ creates an instability drive stronger than the stabilizing influence of magnetic curvature \cite{Drake1980}. 
In figure~\ref{fig:MTM}, the destabilizing effect of $\nabla T_\mathrm{e}$ is illustrated using the thermal drag force $R_\mathrm{T\parallel}\propto(\nu_\mathrm{e}^+-\nu_\mathrm{e}^-)$, where the difference in collision frequency $\nu_\mathrm{e}$ along a field line is due to changes in the electron temperature since $\nu_\mathrm{e}\propto T_\mathrm{e}^{-3/2}$. Importantly, plasma motion at the electron diamagnetic velocity $v_\mathrm{e*}$ introduces a time-lag to $R_\mathrm{T\parallel}$. As a result, the emergent parallel electric field $E_\mathrm{\parallel}$ creates an inductive field that adds in-phase to the initial perturbation, leading to growth of the instability. 

This description brings to light two important facets of MTM instability drive: (1) MTMs should be localized around rational magnetic surfaces and (2) MTMs should oscillate at the electron diamagnetic frequency $\omega_\mathrm{e*}(\psi_\mathrm{n})$, where $\psi_\mathrm{n}$ is a radial unit given by the normalized poloidal flux,. Here $\omega_\mathrm{e*}$ is given by
\begin{equation}
    \label{eq:omega_*e}
    \omega_\mathrm{e*} = k_\mathrm{y} \rho_\mathrm{s} c_\mathrm{s} \bigg(\frac{1}{L_{n_\mathrm{e}}} + \frac{1}{L_{T_\mathrm{e}}}\bigg),
\end{equation}
which depends explicitly on the density and temperature gradient length scales $L_{n_\mathrm{e}}$ and $L_{T_\mathrm{e}}$. More details are given in the methods section. Since $\omega_\mathrm{e*}$ is inversely related to $L_{T_\mathrm{e}}$, a peak in the $\omega_\mathrm{e*}(\psi_\mathrm{n})$ profile corresponds to a peak in the MTM instability drive from $\nabla T_\mathrm{e}(\psi_\mathrm{n})$. Therefore MTMs are most likely to occur when a rational $q$ surface aligns with the peak of the $\omega_\mathrm{e*}(\psi_\mathrm{n})$ profile \cite{Hatch2020}. This formulation has been used to explain steady-state frequency bands observed in magnetic fluctuation data on the JET tokamak, which were identified as MTMs through comparisons with gyrokinetic simulations \cite{Hatch2020}, and it forms the theoretical foundation of the dynamical experimental analysis presented here. 

In plasma experiments, magnetic fluctuations measured in the lab frame will have an additional frequency component given by the Doppler shift $\omega_\mathrm{dop}(\psi_\mathrm{n})$. By exploiting high spatial and temporal resolution diagnostics on the DIII-D tokamak \cite{Eldon2012, Chrystal2016}, we can track the structure of both $\omega_\mathrm{e*}$ and $\omega_\mathrm{dop}$ through time, enabling an investigation of the dynamical evolution of plasma microinstabilities in tokamaks. 

In figure~\ref{fig:normal}, we demonstrate this process for a single $n=3$ MTM, providing unambiguous evidence for MTM activity in the H-mode pedestal. Figure~\ref{fig:normal-a} shows the edge $f_\mathrm{e*,n=3}$ and $f_\mathrm{dop,n=3}$ profiles for a single representative timeslice. As a result of the steep temperature gradients in the pedestal, a large peak in the $n=3$ MTM destabilization potential occurs near the plasma edge. Also shown are the locations of four possible rational $q$ surfaces in the pedestal, with $m$ varying from $15-18$ throughout the steep gradient region. 

In figure~\ref{fig:normal-b}, the evolution of the projected MTM frequency ($f_\mathrm{MTM}=f_\mathrm{e*}+f_\mathrm{dop}$) at these four radial locations is tracked through time between explosive ELM events. The rational surface at $q=16/3$ has the highest MTM instability drive and demonstrates a unique up-chirping frequency behavior after an ELM. Remarkably, this profile-based calculation exactly matches the mode chirping behavior seen in fast magnetic fluctuation measurements, as shown in figure~\ref{fig:normal-c}. Through this theoretically-motivated analysis, we thus explain the distinctive up-chirping behavior observed in magnetic spectrograms\cite{Perez2004,Diallo2015,Laggner2019,Hatch2020,Diallo2021} as follows: The recovery of density and temperature gradients after an ELM \cite{Laggner2019} introduces periodic growth into the $\omega_\mathrm{e*}$ profile described by equation~\ref{eq:omega_*e}. MTMs, being locked at a particular rational $q$ surface, will simultaneously experience a local increase in $\nabla T_\mathrm{e}$ and $\omega_\mathrm{e*}$. Therefore, once these modes turn on at a critical $\nabla T_\mathrm{e}$ \cite{Diallo2015}, their frequency will continue to increase until saturation of the pedestal gradients is achieved. 


\section*{Experimental MTM frequency modification}

\begin{figure*}
    \includegraphics[width=1\linewidth]{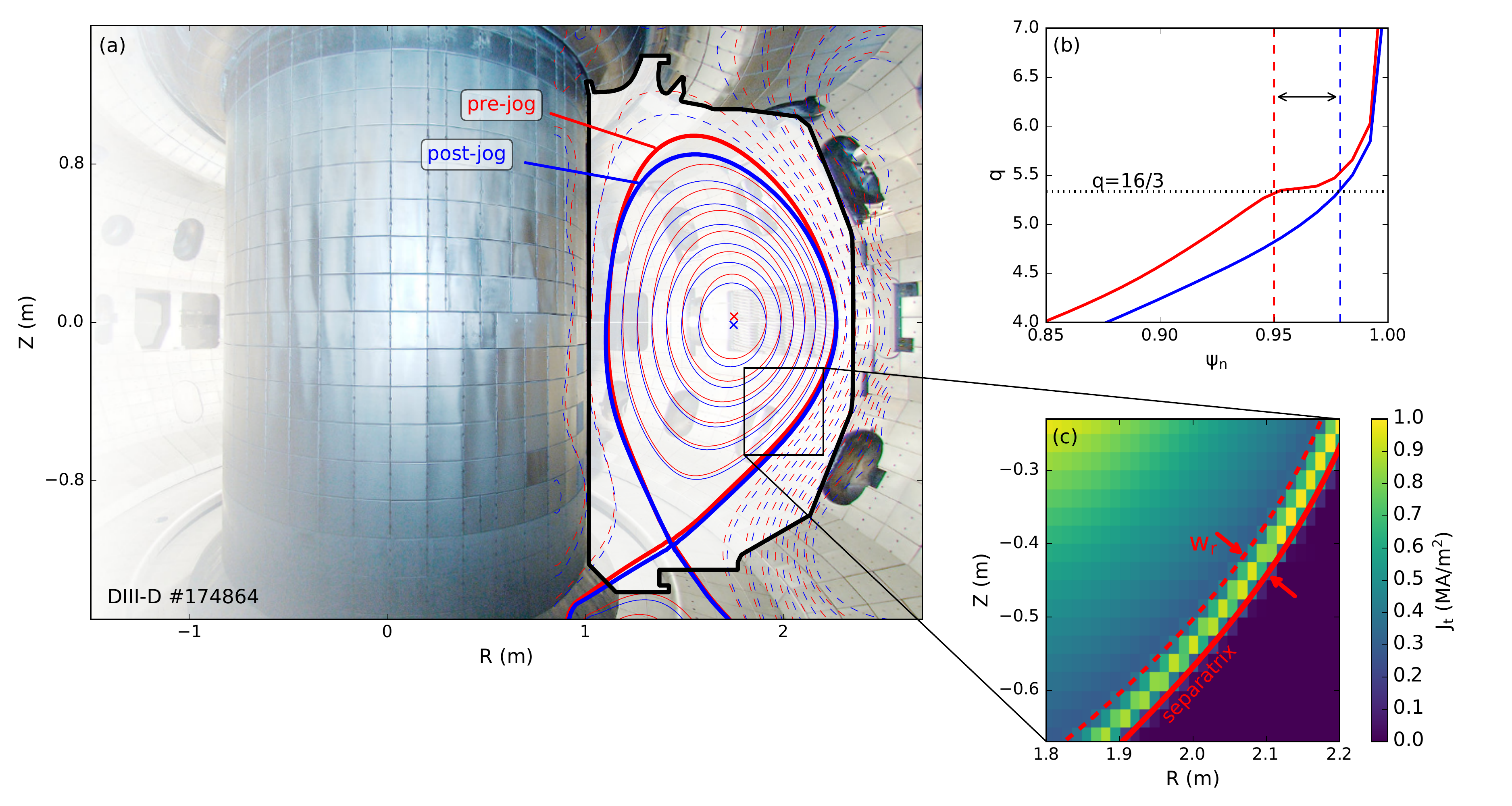}
    \labelphantom{fig:eqs-a}
	\labelphantom{fig:eqs-b}
	\labelphantom{fig:eqs-c}
    \caption{(a) With vertical control algorithms, the plasma is rapidly dropped $\sim10\,$cm during a jogging event. (b) Due to the jog, the resonant $q$ surface moves substantially through the pedestal. (c) The effects of the jog are confined to a small edge region with width $w_\mathrm{r}$ that contains a thin, strong current layer. Image Credit: General Atomics.}
    \label{fig:eqs}
\end{figure*}

\begin{figure}
    \includegraphics[width=1\linewidth]{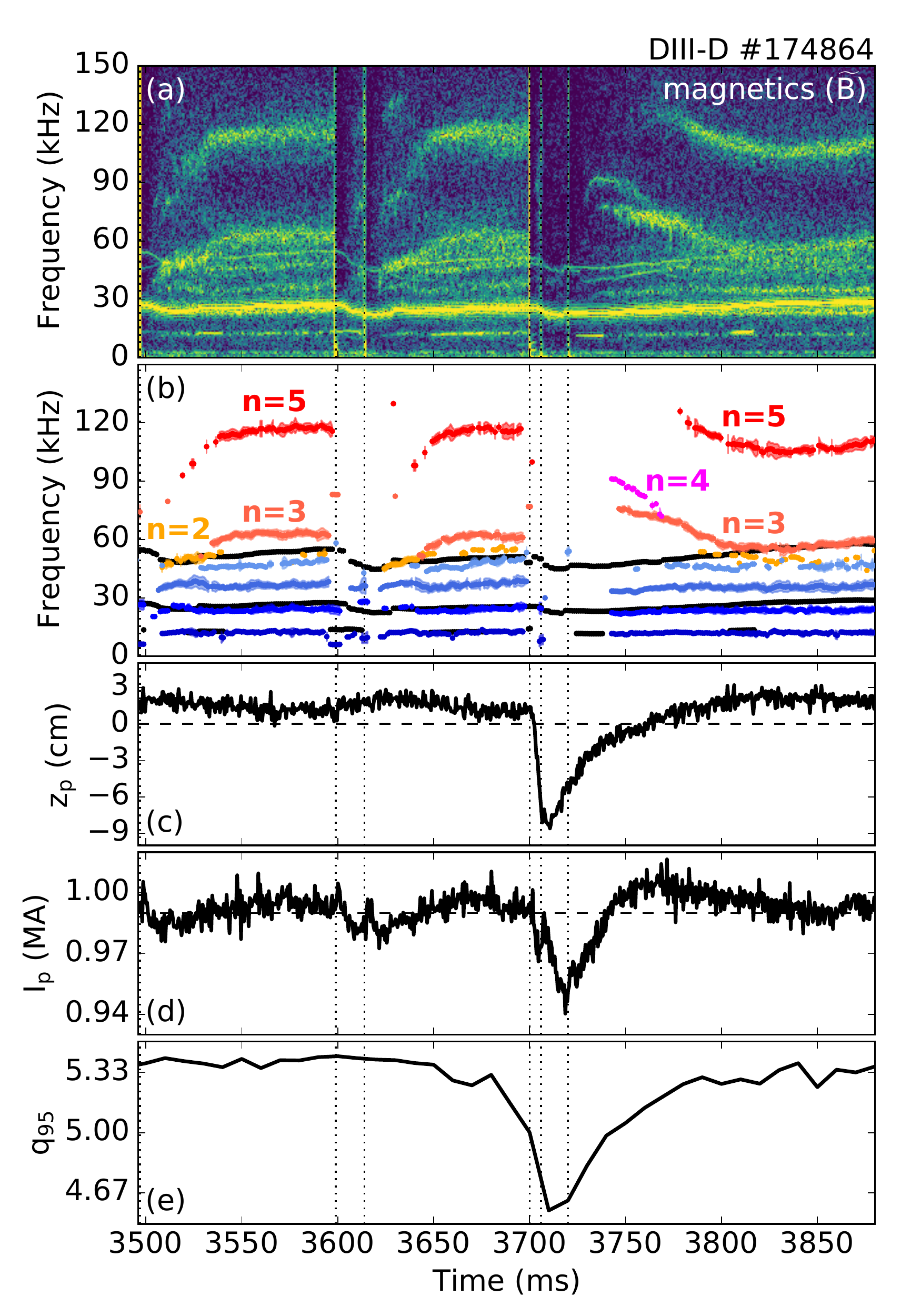}
    \labelphantom{fig:overview-a}
	\labelphantom{fig:overview-b}
	\labelphantom{fig:overview-c}
	\labelphantom{fig:overview-d}
	\labelphantom{fig:overview-e}
    \caption{(a) Magnetic fluctuations show an inverting chirping behavior after a jogging event compared to after a normal ELM. ELM times are indicated with vertical dashed lines. (b) The chirping modes are $n=3, 4$ and $n=5$ modes, shown in orange, magenta and red, respectively. Additional core (black) and pedestal (blue) modes are also shown. The evolution of the (c) magnetic axis $Z_\mathrm{p}$, (d) plasma current $I_\mathrm{p}$ and (e) edge safety factor $q_\mathrm{95}$ through a jogging event show the direct effects of the jog.}
    \label{fig:overview}
\end{figure}

\begin{figure*}
    \includegraphics[width=1\linewidth]{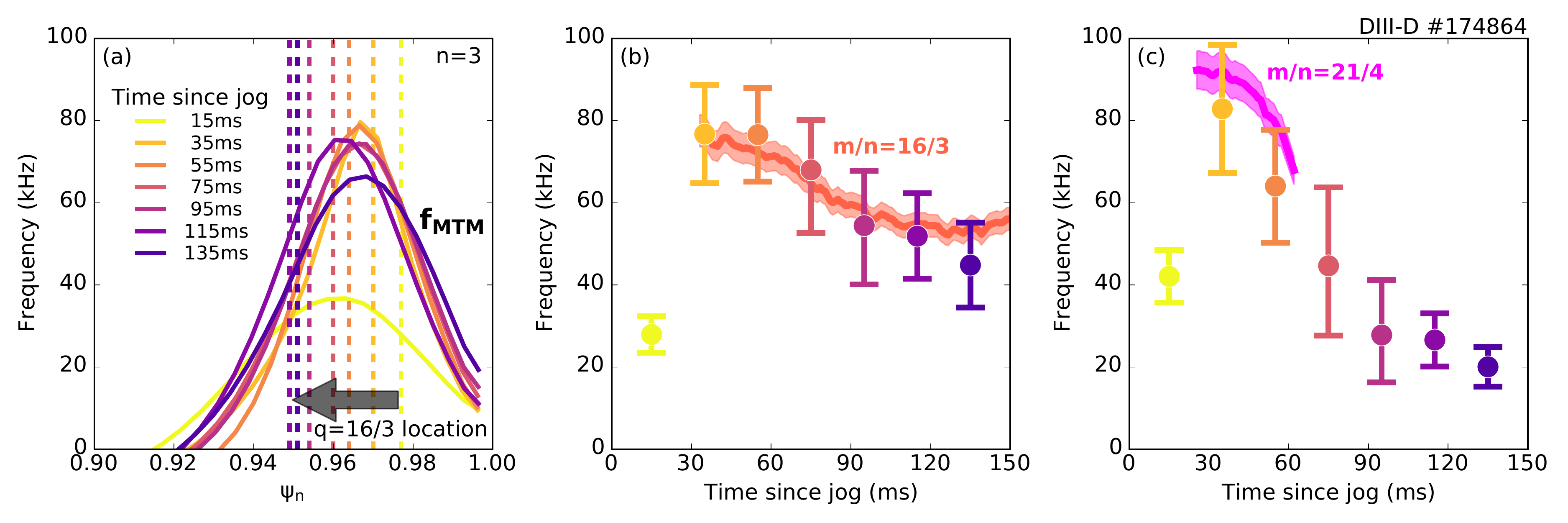}
    \labelphantom{fig:jog-a}
	\labelphantom{fig:jog-b}
	\labelphantom{fig:jog-c}
    \caption{The automated analysis presented in figure~\ref{fig:normal} is repeated for data averaged over several jogging cycles. (a) Due to the current recovery after a jog, the $q=16/3$ surface shifts inwards as pedestal profiles recover.  Incorporating all effects yields excellent agreement between the profile-based $f_\mathrm{MTM}$ predictions and the (b) orange $n=3$ and (c) magenta $n=4$ modes extracted from magnetic fluctuation measurements.}
    \label{fig:jog}
\end{figure*}

With the dynamics of MTM evolution established, we now introduce a novel perturbation scheme to explore the effect of rational surface displacement on the modes. Previously, small vertical oscillations of the plasma volume (``jogs") have been used to perturb the edge current in order to destabilise peeling modes and trigger ELMs \cite{Degeling2003, Lang2004, Gerhardt2010, Kim2012, DeLaLuna2016, Wu2017}. Analytical studies have shown that, during a jogging event, toroidal current is induced in the pedestal due primarily to the compression of the plasma cross section as it travels through an inhomogeneous magnetic field \cite{Artola2018}, as described further in the methods section. Changes in the edge current impact the poloidal magnetic field through Ampere's law, which in turn impacts the winding ratio of magnetic field lines $q=m/n$ and modifies the location of rational magnetic surfaces in the pedestal. 

Here we apply this same principle in a more intense manner with large ($\sim10\,$cm) and fast ($<10\,$ms) downward jogs designed to produce the largest possible perturbations in the edge current. Figure~\ref{fig:eqs-a} compares two equilibria before (red) and after (blue) a large jogging event. The effects of the jog on the plasma are primarily constrained to an edge region $w_\mathrm{r}$, which can be approximated as the MHD skin depth $w_\mathrm{r}\propto\sqrt{\eta}$, where $\eta$ is the plasma resistivity \cite{Artola2018}. As a result of the jog, the reconstructed $q$ profile presented in figure~\ref{fig:eqs-b} dramatically changes. As expected, the radial location of the $q=16/3$ resonant surface moves dramatically through the edge region as an effect of the jog. In figure~\ref{fig:eqs-c}, we show a 2D reconstruction of the plasma current density $J_\mathrm{t}$ for the pre-jog equilibrium, highlighting that the jogs are large enough to influence the edge peak in $J_\mathrm{t}$ but small enough not to significantly disturb the core plasma. This perturbation scheme is uniquely capable of investigating the behavior of microinstabilities in the edge by decoupling the $q$ and $\omega_\mathrm{e*}$ profiles. 

When applied in experiment, the jogs successfully produce clear and quantifiable differences in microinstability signatures distinct from observations during natural inter-ELM periods. In figure~\ref{fig:overview-a}, a magnetic spectrogram from high-frequency Mirnov coils is shown for a time period including two natural ELM periods followed by a large jogging perturbation. Multiple instabilities are evident in the inter-ELM periods, but the higher frequency modes at $\sim60\,$kHz and $\sim110\,$kHz show an inverted chirping behavior after the jogging event at $3700\,$ms. Again, local density fluctuation measurements place these modes in the plasma edge. Using Fourier analysis techniques on a set of fast magnetic diagnostics \cite{Strait2006}, the chirped modes are identified in figure~\ref{fig:overview-b} as $n=3, 4$ and $5$ modes. For comparison, the time dependence of the magnetic axis height ($z_\mathrm{p}$), the total plasma current $I_\mathrm{p}$ and the edge $q$ profile magnitude ($q_\mathrm{95}$) are shown during a jogging event in figure~\ref{fig:overview-c}-(e). 

Notably, the robust analysis developed above can be applied after a jogging event. Figure~\ref{fig:jog-a} shows the edge $f_\mathrm{MTM,16/3}$ profile (solid curves) and the $q=16/3$ location (dashed lines) as a function of post-jog time. During the current recovery period ($\lesssim100\,$ms), the location of the rational $q=16/3$ surface evolves in tandem with the evolution of the profile gradients after a jog-induced ELM. The effect of this motion is that the $q=16/3$ location starts past the peak of the edge $f_\mathrm{MTM}$ profile and then moves inwards over the course of $\sim80\,$ms, falling slightly off the peak destabilizing frequency. 
In contrast to the growth of $f_\mathrm{MTM}$ after a natural ELM, however, the expected MTM frequency falls after a large jogging event due to the inwards motion of the magnetic surface. In figure~\ref{fig:jog-b}, the computed decrease of $f_\mathrm{MTM,16/3}$ is overlayed on the $n=3$ mode extracted from magnetics measurements in figure~\ref{fig:overview-b}. The time-dependent profile analysis matches the experimentally-observed fluctuation dynamics, showing strong agreement between theory and experiment. Moreover, it is noted that the $n=3$ mode amplitude (see figure~\ref{fig:overview-a}) is strongest when the $16/3$ surface is best aligned with the peak of the $\omega_\mathrm{e*}$ profile, in agreement with the expectation that the electron thermal gradient, which peaks with $\omega_\mathrm{e*}$, acts as main MTM drive \cite{Hatch2020}. 

In figure~\ref{fig:jog-c}, the analysis is augmented by matching the decreasing $n=4$ mode observed early in the ELM cycle after jogs. In this case, the rational surface $m/n=21/4$, which lies just inside of the $m/n=16/3$ surface, shows the best alignment with the peak of the $\omega_\mathrm{e*}$ profile. Directly after a jog, the MTM drive on this surface is high and the mode appears in magnetic fluctuation measurements. However, as the $q$ surface moves inwards, the drive, amplitude and frequency drop until the $n=3$ mode dominates. Power is transferred to more unstable rational surfaces during this transition through non-linear coupling between various pedestal modes \cite{Diallo2018}. This manifests in the magnetics measurements as a disappearance of the $n=4$ signature coincident with a peak in the $n=3$ amplitude around $\sim60\,$ms after the jogging event (see figure~\ref{fig:overview-a}). Again, figure~\ref{fig:jog-c} shows excellent agreement between the profile and fluctuations measurements, verifying the dynamic behavior of edge-localized MTMs. We note further that the very low MTM frequency predicted directly after the jog ($t_\mathrm{jog} = 20\,$ms) in figures~\ref{fig:jog-b} and~\ref{fig:jog-c} occurs at a time before the MTM onset and thus is not expected to produce magnetic fluctuations. This analysis can be reproduced to the same effect for the $n=5$ modes shown in figure~\ref{fig:overview-b} and is robust to changes in the background plasma across several DIII-D discharges, showing that minor background plasma changes do not modify the fundamental MTM behavior.

\section*{Discussion and Outlook}

To support the above dynamic identification of MTMs in the H-mode pedestal, we note here that additional measurements were taken to rule out the possibility of the above dynamics being caused by other instabilities. 
\begin{itemize}
    \item Large amplitudes of these modes in magnetic fluctuation measurements suggest that the modes are electromagnetic in nature, eliminating consideration of ETGs, ITGs and TEMs.
    \item The propagation of the chirped modes is determined to be strongly in the electron diamagnetic direction, as is expected for MTMs, both through $E\times B$ profile calculations and detailed analysis of local density fluctuation data \cite{McKee1999}. Conversely, KBMs (the other primary electromagnetic edge turbulence candidate) rotate in the ion diamagnetic direction, which is inconsistent with the measured data. 
    \item Experimental transport studies show that $D_\mathrm{e}/\chi_\mathrm{e}\sim1/10$ in the vicinity of the $n=3$ and $n=5$ chirped modes. This is consistent with gyrokinetic predictions stating that pedestal MTMs should contribute predominately to electron thermal transport, whereas KBMs contribute approximately equally to both electron thermal and particle transport \cite{Kotschenreuther2019}. 
    \item Finally, the modes turn on simultaneously with the saturation of $\nabla T_\mathrm{e}$, consistent with models tying destabilization to the growth of pedestal gradients beyond a critical threshold \cite{Diallo2014, Diallo2015}. 
\end{itemize} 
In sum, these compounding observations combined with the novel dynamic frequency evolution described above paint an explicit experimental picture of the existence of MTMs in the DIII-D H-mode pedestal. Further, measurements showing that the associated transport is predominantly in the electron heat channel and is closely linked to the saturation of $\nabla T_\mathrm{e}$ reinforce the established tendency of MTMs to contribute significantly to electron heat flux through the plasma edge. 

By regulating electron thermal transport, MTMs are expected to establish limits on the maximum electron temperature gradients within the H-mode pedestal \cite{Dickinson2012}. Since the pedestal contributes substantially to total plasma pressure and fusion performance, identifying and understanding the full impact of MTMs could have significant implications on the design of future pilot plan scenarios.  
While numerous theoretical studies have predicted the presence of these instabilities under common pedestal conditions, an experimental validation is required to build confidence in the MTM physics basis if predictive pedestal models are to be expanded to include the relevant effects. In this article, we experimentally demonstrate the existence of edge-localized MTMs by exploiting the time response of the plasma to vertical jogs.  
These results validate leading analytic and numerical theories \cite{Kotschenreuther2019, Hatch2020} and motivate the future incorporation of MTM effects into advanced predictions of the full plasma performance. 

The success of the perturbative approach applied here also implies a possibility for the expansion of dynamic turbulence identification to other unexplained tokamak regimes. Similar instability markers have been reported on a wide variety of machines and scenarios \cite{Laggner2019, Diallo2021}, but the underlying physics remains largely undetermined. The analysis presented here offers a new mechanism to uncover explanations for these observations, potentially enabling a comprehensive perturbative study of experimental transport signatures in tokamak devices. 


\section*{Methods}

This investigation is predicated on the accurate simultaneous measurement of many different plasma parameters, including electron and ion densities and temperatures and the equilibrium magnetic structure. The Doppler-shift $\omega_\mathrm{dop}$ and electron diamagnetic frequency $\omega_\mathrm{*e}$ are calculated from Carbon impurity measurements from charge-exchange recombination \cite{Chrystal2016} and electron profile measurements from Thomson scattering \cite{Eldon2012}, respectively. Localized density fluctuations measurements are made with beam emissions spectroscopy in order to localize the instabilities in the plasma edge \cite{McKee1999}. To acquire robust statistics, measurements are taken every $\sim20\,$ms throughout several dedicated discharges (each of which lasts $\sim5\,$s) and then reordered on a single timebase defined to illuminate reproducible profile evolution between periodic ELM events, as is plotted in the figures. This reduces statistical noise caused by slight variation in plasma parameters over time. 

All data in this study are mapped on to detailed kinetic equilibrium reconstructions, which provide the magnetic field distribution throughout the plasma, in order to best capture the edge dynamics \cite{Xing2021}. Essential to these reconstructions are accurate calculations of the bootstrap current \cite{Bickerton1971}, requiring the consideration of constraints from both magnetic and internal profile data. Senstive equilibrium reconstructions are necessary to calculate both the edge $q$ profile, for which no direct measurement currently exists on DIII-D, and the radial alignment between the $\omega_\mathrm{dop}$ and $\omega_\mathrm{*e}$ profiles. To facilitate robust analysis, multiple reconstructions are made for each measurement time to generate effective uncertainties in the plasma magnetic structure, which are propagated through the final MTM frequency calculation.

Throughout this work, we define $\omega_\mathrm{e*}$ as in equation~\ref{eq:omega_*e}, reproduced here for convenience: 
\begin{equation}
    \omega_\mathrm{e*} = k_\mathrm{y} \rho_\mathrm{s} c_\mathrm{s} \bigg(\frac{1}{L_{n_\mathrm{e}}} + \frac{1}{L_{T_\mathrm{e}}}\bigg).
\end{equation}
Here $k_y=nq/a\rho_\mathrm{tor}$ is the binormal wavenumber, $\rho_\mathrm{tor} = \sqrt{\Phi_n}$ is the square root of the normalized toroidal magnetic flux, $\rho_\mathrm{s}=c_\mathrm{s}/\Omega_\mathrm{i}$ is the sound gyroradius, $c_\mathrm{s}=\sqrt{ZT_\mathrm{e}/m_i}$ is the sound speed, $\Omega_\mathrm{i}$ is the ion gyrofrequency, and the electron density and temperature gradient scale lengths are defined as $a/L_{n_\mathrm{e}} = (1/n_\mathrm{e}) (dn_\mathrm{e}/d\rho_{tor})$ and $a/L_{T_\mathrm{e}} = (1/T_\mathrm{e}) (dT_\mathrm{e}/d\rho_{tor})$, respectively \cite{Hatch2020}. The Doppler shift $\omega_\mathrm{dop}$ is correspondingly given by:
\begin{equation}
    \label{eq:omega_dop}
    \omega_\mathrm{dop} = \frac{nE_\mathrm{r}}{RB_\mathrm{p}},
\end{equation}
where $n$ is the toroidal mode number, $E_\mathrm{r}$ is the radial electric field, $R$ is the major radius and $B_\mathrm{p}$ is the poloidal magnetic field. 

During a jogging event, the induction of current due to the motion of the plasma through an inhomogeneous magnetic field can be by is described by
\begin{equation}
    \delta I_\phi^{w_\mathrm{r}} = \frac{4\pi}{\mu_0R_0} \bigg[ \delta\psi_\mathrm{ext}(a) - B_\theta(r_0)R_0\delta w_\mathrm{r} - \eta J_\phi \delta t\bigg],
    \label{eq:Artola}
\end{equation}
as first reported by Artola \textit{et. al.}\cite{Artola2018}. Here the change in total toroidal edge current ($\delta I_\phi^{w_\mathrm{r}}$) is given as a function of the local change in external magnetic flux ($\delta\psi_\mathrm{ext}$), the inhomogeneous poloidal magnetic field ($B_\theta$), plasma compression ($\delta w_\mathrm{r}$) and a small resistive decay term ($\eta J_\phi \delta t$) \cite{Artola2018}. The width $w_\mathrm{r}$ of the edge region is generally small compared to the plasma minor radius ($r_\mathrm{0}$) such that $w_\mathrm{r}/r_0<<1$ and can be approximated as the skin depth $w_\mathrm{r}\sim\sqrt{\eta/(\pi\mu_0f)}$, where $f$ is the oscillation frequency and $\eta$ is the plasma resistivity.

During and after a jogging event, the corresponding changes in the current profile are taken from kinetic equilibrium reconstructions based on fast magnetic measurements and internal plasma profiles. In tokamaks, the safety factor $q=m/n$ can be defined as
\begin{equation}
    \label{eq:q}
    q=\frac{rB_\phi}{RB_\theta},
\end{equation} where $r$ is the minor radius, $R$ is the major radius, and $B_\phi$ and $B_\theta$ are the toroidal and poloidal magnetic fields, respectively. Since $B_\theta$ is directly related to the toroidal plasma current through Ampere's Law, the $q$ profile in the plasma edge is significantly modified during a jog. This is the necessary perturbation for the study of edge microinstabilities, as is discussed in the main text. 

Throughout this work, 1D and 2D transport simulations are conducted with the TRANSP \cite{Poli2018} and autoUEDGE \cite{Izacard2018, Nelson_UEDGE} codes in order to verify the transport fingerprints of pedestal MTMs. Assessment of the toroidal mode numbers was completed with the MODESPEC code; magnetic diagnostic resolution is not high enough to directly determine the poloidal mode numbers \cite{Strait2006}, suggesting $m>12$ as found through the profile analysis. Part of data analysis for this work was performed using the OMFIT integrated modeling framework \cite{Meneghini2015, Logan2018}.

\section*{Data availability}
The data discussed and used for all figures in this article are available from the corresponding author upon reasonable request.

\section*{acknowledgments}
The authors would like to especially thank D.R. Hatch and M. Curie for several helpful discussions relating to the frequency identification of MTMs, as well as W. Guttenfelder and R. Nazikian for valuable advice during preparation of the manuscript. This material was supported by the U.S. Department of Energy, Office of Science, Office of Fusion Energy Sciences, using the DIII-D National Fusion Facility, a DOE Office of Science user facility, under Awards DC-AC02-09CH11466, DE-SC0015480, DE-SC0015878 and DE-FC02-04ER54698. This report is prepared as an account of work sponsored by an agency of the United States Government. Neither the United States Government nor any agency thereof, nor any of their employees, makes any warranty, express or implied, or assumes any legal liability or responsibility for the accuracy, completeness, or usefulness of any information, apparatus, product, or process disclosed, or represents that its use would not infringe privately owned rights. Reference herein to any specific commercial product, process, or service by trade name, trademark, manufacturer, or otherwise, does not necessarily constitute or imply its endorsement, recommendation, or favoring by the United States Government or any agency thereof. The views and opinions of authors expressed herein do not necessarily state or reflect those of the United States Government or any agency thereof.

\bibliographystyle{apsrev4-2}
\bibliography{main} 

\end{document}